\documentclass[twocolumn,prb,showpacs,multicol,amsmath,amssymb]{revtex4-1}
\usepackage{amssymb}
\usepackage{amsmath}
\usepackage{graphicx}
\usepackage{epstopdf}
\usepackage{bm}
\usepackage{gensymb}

\newcommand{\be}{\begin{equation}}
\newcommand{\ee}{\end{equation}}
\newcommand{\bk}{{{\bf{k}}}}

\newcommand{\bea}{\begin{eqnarray}}
\newcommand{\eea}{\end{eqnarray}}

\newcommand{\nn}{\nonumber}

\newcommand{\bd}{\begin{displaymath}}
\newcommand{\ed}{\end{displaymath}}
\newcommand{\ba}{\begin{array}}
\newcommand{\ea}{\end{array}}
\newcommand{\bi}{\begin{itemize}}
\newcommand{\ei}{\end{itemize}}
\newcommand{\bc}{\begin{center}}
\newcommand{\ec}{\end{center}}
\newcommand{\bfl}{\begin{flushleft}}
\newcommand{\efl}{\end{flushleft}}
\newcommand{\bfr}{\begin{flushright}}
\newcommand{\efr}{\end{flushright}}

\def\bk{{\bf k}}   
  \def\bd{{\bf d}}

\def\6{\partial}

\def\o{\omega}

\def\={\!\!\!&=&\!\!\!}
\def\+{\!\!\!&&\!\!\!+~}
\def\-{\!\!\!&&\!\!\!-~}

\newcommand\redout{\bgroup\markoverwith{\textcolor{red}{\rule[.5ex]{2pt}{0.4pt}}}\ULon}
\usepackage[colorlinks=true,citecolor=blue]{hyperref}
\hypersetup{colorlinks=true,citecolor=blue,linkcolor=red,urlcolor=blue}

\newcommand{\Nb}{NbSe$_2$}
\graphicspath{{.}{./plot/}}

\begin{document}

\title{Proximity-Induced Mixed Odd and Even Frequency Pairings in Monolayer \Nb~}

\author{Mojtaba Rahimi Aliabad}\email{mrahimi93@gmail.com}
\affiliation{Nano Structured Coatings Institute of Yazd Payame Noor University, PO Code: 89431-74559, Yazd, Iran}
\author{Mohammad-Hossein Zare}\email{zare@qut.ac.ir}
\affiliation{Department of Physics, Faculty of Science, Qom University of Technology, Qom 37181-46645, Iran}


\date{\today}


\begin{abstract}
Monolayer superconducting transition metal dichalcogenide \Nb~is a candidate for a nodal 
topological superconductor by magnetic field.
Because of the so-called Ising spin-orbit coupling that strongly pins the electron spins to the out-of-plane direction,
Cooper pairs in monolayer superconductor \Nb~are protected against an applied in-plane
magnetic field much larger than the Pauli limit.
In monolayer \Nb, in addition to the Fermi pockets at the corners
of Brillouin zone with opposite crystal momentum similar to other semiconducting transition metal dichalcogenids,
there is an extra Fermi pocket around the $\Gamma$ point with
much smaller spin splitting, which could lead to an alternative strategy for pairing possibilities that are manipulable by
a smaller magnetic field.
By considering a monolayer \Nb-ferromagnet substrate junction,
we explore the modified pairing correlations on the pocket at $\Gamma$ point in hole-doped monolayer \Nb.
The underlying physics is fascinating as there is a delicate interplay of the induced exchange field and the Ising spin-orbit coupling.
We realize a mixed singlet-triplet superconductivity, $s+f$, due to the Ising spin-orbit coupling.
Moreover, our results reveal the admixture state including both odd- and even-frequency components, associated with the ferromagnetic proximity effect.
Different frequency symmetries of the induced pairing correlations can be realized by manipulating the magnitude and direction of the induced magnetization.        
\end{abstract}
\maketitle
\section{Introduction}
The search for unconventional superconductors (SC) such as high-temperature SC~\cite{Lee:2006,Keimer:2014}, noncentrosymmetric SC~\cite{Bauer:2012,Yip:2014,Smidman:2017} and
topological SC~\cite{Beenakker:2013,Ando:2015,Masatoshi:2017} has witnessed growing interest during the past decade. 
In noncentrosymmetric SC,
an anisotropic spin-orbit coupling (SOC) is created in these materials because they lack a center of inversion.
Based on the amplitude of the anisotropic SOC, the noncentrosymmetric materials exhibit a number of interesting features,
in contrast to the centrosymmetric materials.
For example, the usual classification of the superconducting pairings in terms of their spatial symmetry
is no longer valid in these systems because the parity is not a good quantum number.
As a result, lack of inversion symmetry supports the pairing wave function including both parity-even and parity-odd functions~\cite{Bauer:2012}. 

Unconventional odd-frequency superconductivity is a dynamical phenomenon in which the fermionic wave function of the Cooper pair
has to change sign when interchanging two particles in time or equivalently be odd in frequency due to the Pauli exclusion principle.
Unequal time pairings could lead to an extension of the superconducting gap parameters to states such as
spin-triplet even-parity or spin-singlet odd-parity states~\cite{Bergeret:2005,Linder:2017}.
Berezinskii\cite{Berezinskii:1974} proposed odd-frequency pairing for describing the $^3$He superfluidity 
and then later in superconductivity~\cite{Balatsky:1992}.
In recent years, many theoretical studies and experimental indications have demonstrated the possibility of an odd-frequency SC in different heterostructures 
~\cite{Bergeret:2001,Volkov:2003,Linder:2009,Tanaka:2012,Annica:2013,Triola:2014,Bernardo:2015}.
Furthermore, the theoretical investigations have illustrated the existence of the odd-frequency superconducting pairing
in multiorbital superconductors due to the band label behaviors as an additional
symmetry~\cite{Black-Schaffer:2013,Gao:2015,Nomoto:2016,Komendova:2017,Triola:2017} and in two dimensional materials coupled to
superconductors~\cite{Parhizgar:2014,Triola:2016}.
Broken time-reversal symmetry occurs by magnetic field;
it is possible to create an admixture state including both odd- and even-frequency components.
Recent studies indicate the existence of an important connection between odd-frequency pairing correlation and
zero Majorana fermions~\cite{Linder:2017,Tanaka:2012,Asano:2013,Snelder:2015,Huang:2015,Ebisu:2015,Kashuba:2017}.
Since the Majorana fermion operator is Hermitian, i.e., $\gamma^{\dagger}=\gamma$, the resulting
normal Green's-function $G(\tau)=-\langle {T_{\tau}}\gamma^{\dagger}({\tau})\gamma(0)\rangle$ is at the same time the anomalous Green's-functions
${\cal F({\tau})}=-\langle {T_{\tau}}\gamma({\tau})\gamma(0)\rangle$.
Note that a single Majorana bound state has no extra degrees of freedom such as spin, or momentum, and
therefore the odd-frequency state is possible because the time retarded correlator $\langle {T}_{\tau}\gamma({\tau})\gamma(0)\rangle$ is odd in the time argument.
In the case of zero-energy Majorana fermions, $G({\omega_n})={\cal F}({\omega_n})={1}/{{i\omega_n}}$.
Consequently, the Majorana fermions can be utilized as a platform for realizing the odd-frequency pairing correlations.

Monolayer transition-metal dichalcogenides (TMDs), a large family of two-dimensional materials, have a honeycomb lattice structure 
similar to graphene from the top point of view.
TMDs consist of three atomic layers wherein a transition metal layer is sandwiched between two outer chalcogen layers. The atoms are arranged in a triangular
configuration in each layer~\cite{Bromley:1972,Boker:2001}.
Monolayer TMDs have the point group symmetry  $D_{3h}$, lacking inversion symmetry.
Therefore, they are inherently noncentrosymmetric systems.
In the monolayer TMDs, a special type of SOC is produced (also called Ising SOC) due to the strong atomic SOC of the $d$ orbitals of the heavy metal atoms
and the broken in-plane inversion symmetry \cite{Zhu:2011,Xio:2012,Korm:2013,Lu:2015,Xi:2016,Zhou:2016}.  
Ising SOC gives rise to out-of-plane spin polarization in contrast to the Rashba SOC, which polarizes the electron spins in the in-plane direction.
The Ising SOC lifts the spin degeneracy of fermions in momentum space, which could lead to an alternative strategy for pairing possibilities
~\cite{Zhou:2016,Yuan:2014} and spin-polarized currents~\cite{Vorontsov:2008,Tanaka:2009} in TMDs.
More recently, discovering intriguing phenomena such as unconventional superconductivity~\cite{Lu:2015,Shi:2015,Saito:2016,Rahimi:2017},
coexistence of SC and charge density wave~\cite{Xi:2015,Cao:2015,Ugeda:2015} and Bose-metal phase~\cite{Tsen:2015} at the ultrathin films of TMDs
is presently generating much interest.
Experimental studies have demonstrated the existence of an unconventional SC unlike s-wave SC, in single layer TMDs such as \Nb~~\cite{Xi:2016,Barrera:2018}
and MoS$_2$~\cite{Lu:2015,Zhou:2016} in critical upper in-plane magnetic field
several times higher than the Pauli paramagnetic limit. 
Experimental observations show that the Ising SC would protect against an external in-plane magnetic field,
owing to the Ising SOC, which strongly pins electron spins to the out-of-plane direction~\cite{Lu:2015,Saito:2016,Xi:2015}. 
Among the TMD materials, recently studying electronic and SC properties of \Nb~has attracted many attentions. The thickness of the ultrathin films of \Nb,
owing to the relatively weak van der Waals interaction between layers, can be exfoliated down to
the monolayer limit~\cite{Xi:2015,Cao:2015,Tsen:2015,Xi:2016,Frindt:1972,Staley:2009}.
The superconducting transition temperature increases monotonically with increasing layer thickness~\cite{Xi:2015,Cao:2015,Tsen:2015,Xi:2016}. 
Monolayer \Nb~ (ML-NS) has been successfully grown
on bilayer graphene by using molecular beam epitaxy~\cite{Ugeda:2015}. In contrast to other semiconducting TMDs such as MoS$_2$, MoSe$_2$ and WS$_2$
which are superconducting under optimal electro gating~\cite{Lu:2015,Shi:2015,Saito:2016} and under pressure~\cite{Chi:2018},
ML-NS is inherently a conventional $s$-wave type superconductor with a fully-gapped Fermi-surface with a critical temperature about 3K~\cite{Xi:2015,Xi:2016,Cao:2015,Tsen:2015}.

\begin{figure}[t]
 \begin{center}
\vspace{0.10cm}
 \hspace{-0.1cm}
\includegraphics[width=0.44 \linewidth]{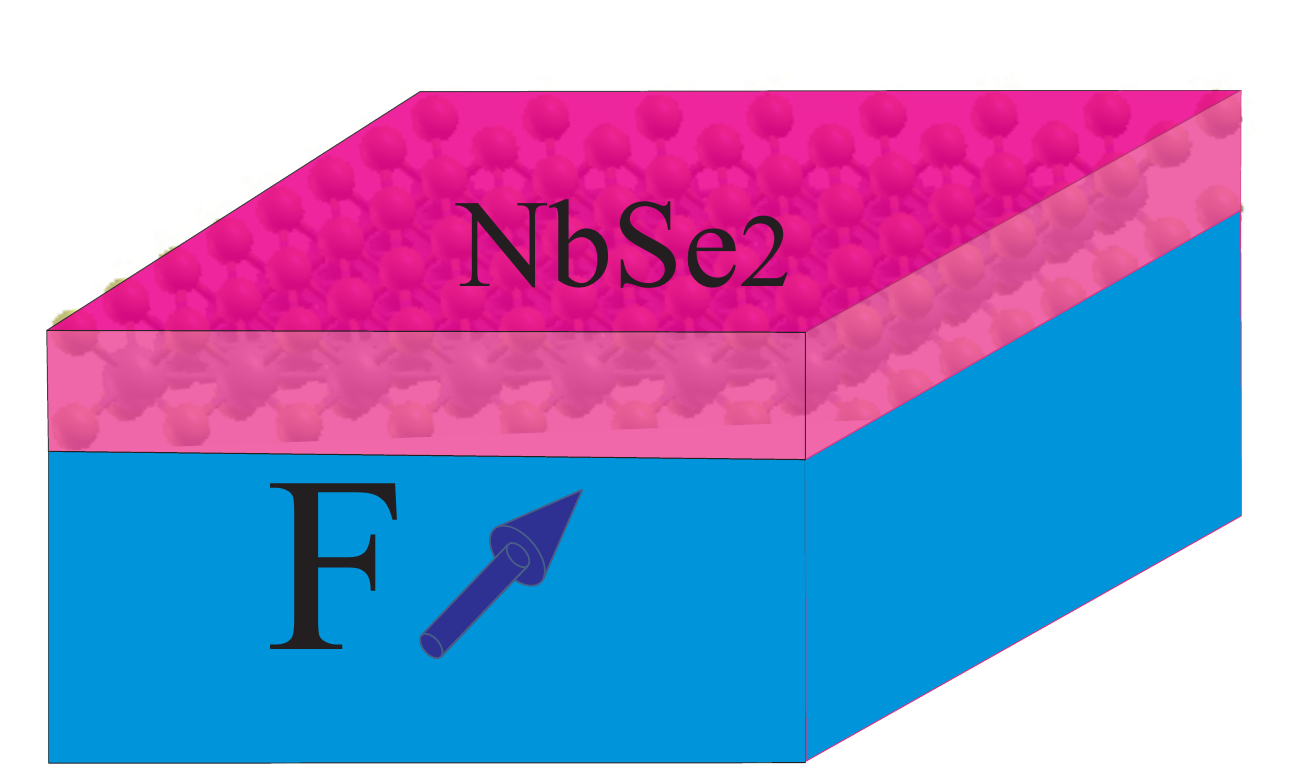}  \hspace{.0cm}
\put (-14.3, .4){\makebox[0.01\linewidth][l]{(a) }}
\includegraphics[width=0.5\linewidth]{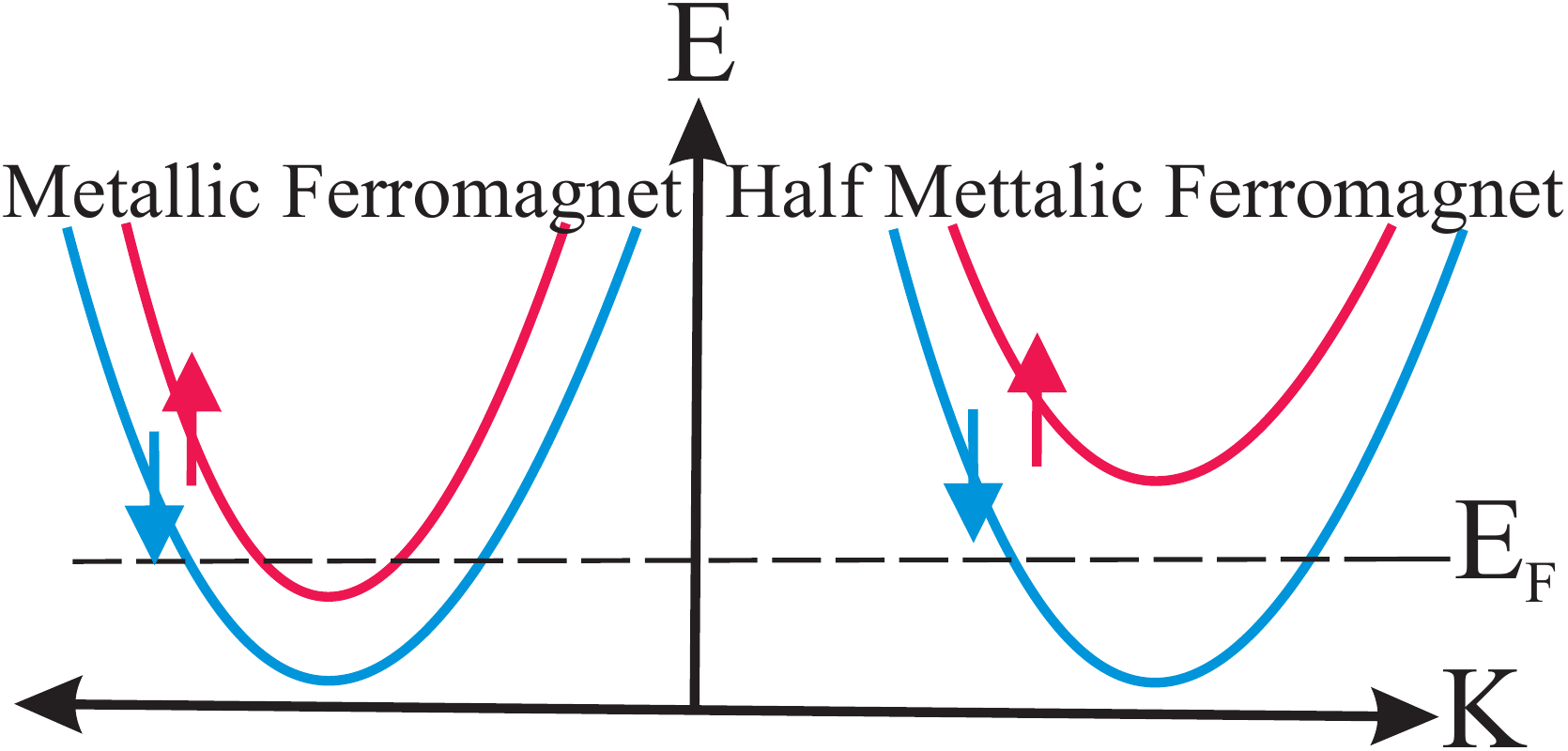} \hspace{.0cm}
\put (-4.5, .4){\makebox[0.01\linewidth][l]{(b) }}\vspace{.7cm}
 \includegraphics[width=0.44\linewidth]{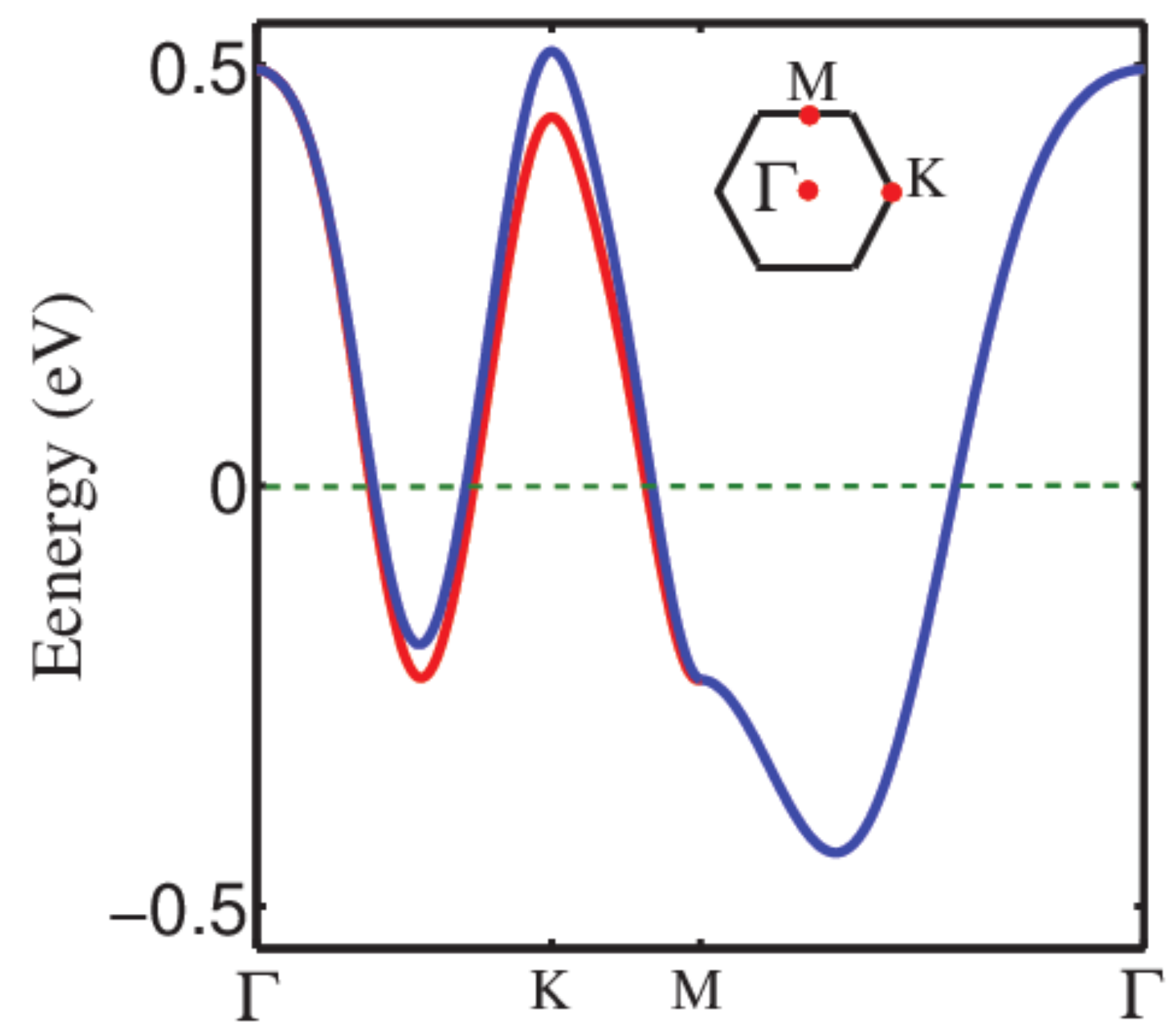}  \hspace{.0cm}
 \put (-4.3,1.1){\makebox[0.01\linewidth][l]{(c) }}
 \includegraphics[width=0.53\linewidth]{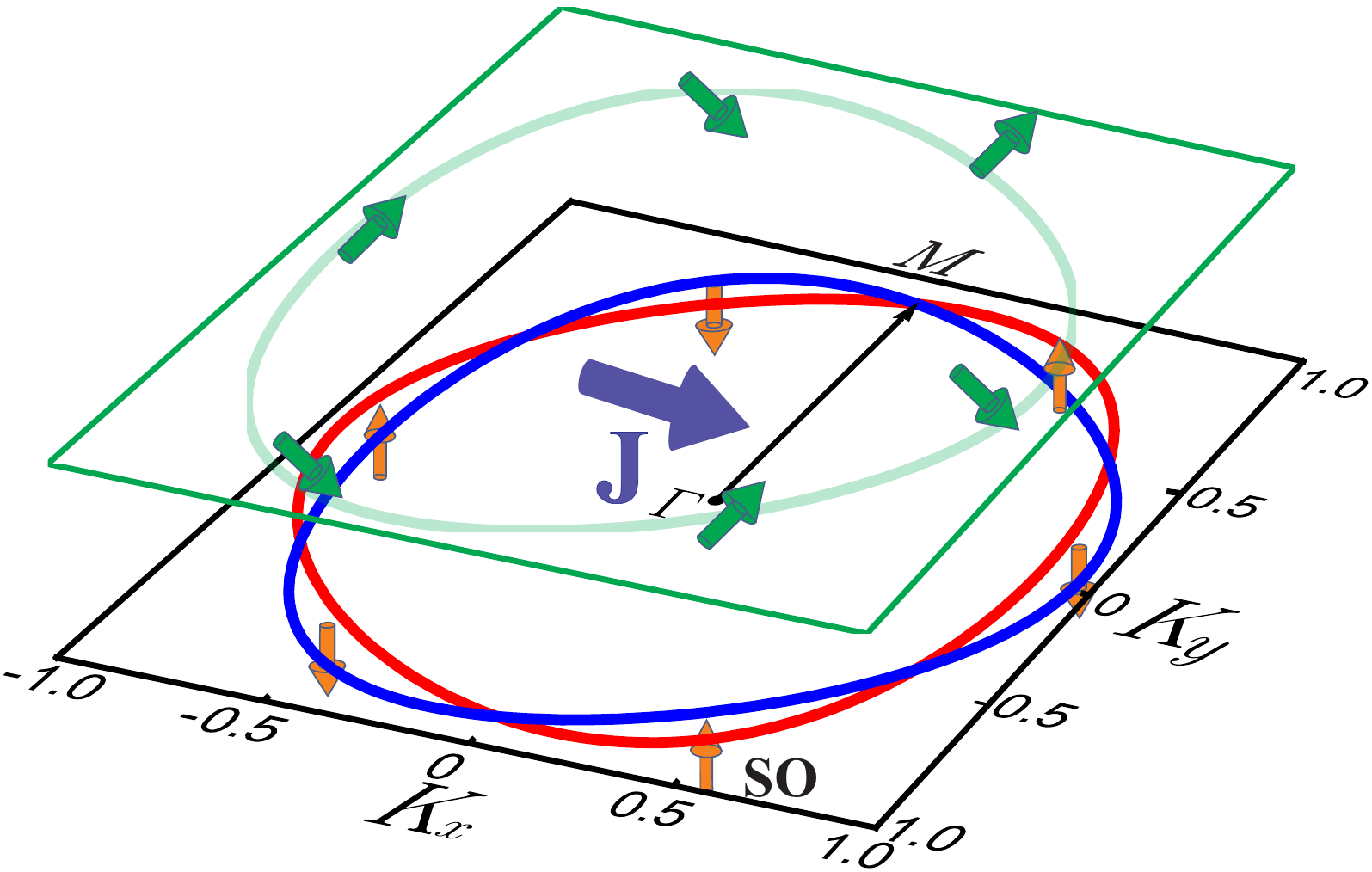}  \hspace{.0cm}
 \put (-4.5, 1.1){\makebox[0.01\linewidth][l]{(d) }}
\end{center}
 \vspace{-0.2cm}
\caption{(Color online) 
(a) Schematic illustration of a ML-NS/ferromagnet heterostructure.
(b) An intuitive understanding of the band structures of a metallic ferromagnet and a half-metallic ferromagnet.
(c) The obtained band structure of ML-NS via a tight-binding calculation. 
(d) Schematic illustration of Fermi surface hosting up- and down-spins with respect to the $\Gamma$ point shown in red and blue contour plots, respectively.
The green contour shows the induced non-collinearity due to the proximity effect of the ferromagnet substrates.
}
\label{fig:1}
\end{figure}

\begin{figure}[t]
\vspace{-0.7cm}
\includegraphics[width=0.95 \linewidth]{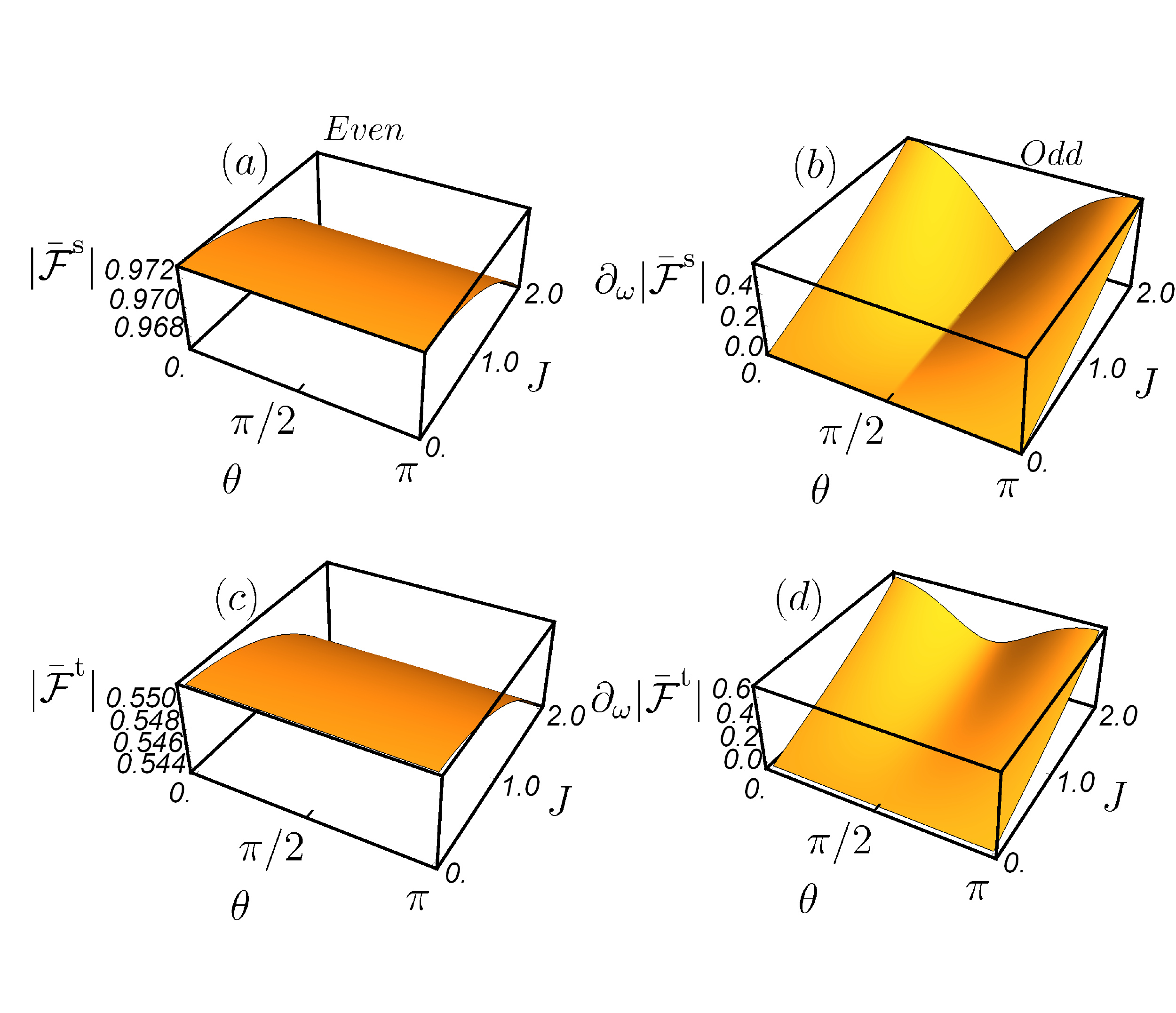}
\vspace{-0.7cm}
\caption{ Magnitude of the even-frequency, ${\hat{\mathcal{F}}}({\bf k},\omega=0)$, and
the odd-frequency superconducting pairing correlations, ${{\partial_{\omega}{\hat{\mathcal{F}}}}({\bf k},\omega)|_{\omega=0}}$, in the plane of $J$ and $\theta$.
Here, $J=\sqrt{J_x^2+J_z^2}$ and $\theta=\arctan(J_x/J_z)$.
The results shown are for $k_x=0.6, k_y=0$ and $\omega=0$, with the same behavior for all frequencies and any ${\bf k}$ as well.
} 
\label{fig:2}
\end{figure}

In the present paper, 
we study proximity-induced superconductivity in ML-NS due to an external exchange field originating from different ferromagnet substrates
by considering a ML-NS/ferromagnet junction as depicted in Fig.~\ref{fig:1}(a).
The relevant band dispersions for the two different magnetic substrates such as metallic ferromagnet and half metallic ferromagnet shown in Fig.~\ref{fig:1}(b).
As is clear, in the half-metallic ferromagnet, only one spin subband is partially filled. 
In the case of the metallic ferromagnet, the chemical potential crosses both spin subbands causing both spin channels to contribute to conduction.   
As will be discussed, as a result of the interplay of Ising SOC and magnetization, various superconducting order parameters appear in ML-NS, which could be symmetric
or antisymmetric with respect to the spin degree of freedom and as well as odd or even in frequency. 
For this end, we compute the modified anomalous correlation function in ML-NS using the Matsubara Green's-function formalism~\cite{Bruus:2004}. 
Then, we proceed to perform a symmetry analysis on the modified pairing potential in ML-NS due to the proximity effect.
Our results reveal that
it is feasible to engineer the frequency symmetry of the superconducting order parameters in ML-NS by varying the magnetization orientation as shown in Fig.~(\ref{fig:2}).
In the absence of the exchange field, the mixed superconducting order parameters are even in frequency,
while the interplay of exchange field and SOC could give rise to the admixture pairing correlations including both even- and odd-frequency components.
Finally, we will continue with a brief discussion of the created pairing correlations in ML-NS
due to the proximity effect of different ferromagnet substrates such as metallic and half metallic ferromagnets.

%

\section{Hamiltonian} 

The six-band tight-binding model recently proposed for ML-NS in the basis of
$\Psi_{\bf k}=(c_{{\bf k},d_{z^2},\uparrow},c_{{\bf k},d_{xy},\uparrow},c_{{\bf k},d_{x^2-y^2},\uparrow},c_{{\bf k},d_{z^2},\downarrow},c_{{\bf k},d_{xy},\downarrow},c_{{\bf k},d_{x^2-y^2},\downarrow})$,
is given by \cite{He:2016}
\begin{align}
 \begin{aligned}
  {\cal H}=\sum_{\bf k} \Psi^{\dagger}_{\bf k} {\cal H} ({\bf k}) \Psi_{\bf k}
  \label{eq:1}
 \end{aligned}
\end{align}
where
\begin{align}
 \begin{aligned}
 &
{\cal H} ({\bf k})= {\cal H}_0 ({\bf k}) \otimes \sigma_0 + \frac{1}{2} \lambda_{so} L_{z}\otimes \sigma_z
\label{eq:1a}
 \end{aligned}
 \end{align}
and where $c_{{\bf k},\alpha}$ $(c^{\dagger}_{{\bf k},\alpha})$ annihilates (creates) an electron with momentum ${\bf k}=(k_x,k_y)$ and spin $\alpha=\uparrow,\downarrow$,
$\lambda_{so}$ is the strength of the intrinsic Ising SOC in ML-NS and
${\sigma_0},~\sigma_z$ are the identity and Pauli matrices, respectively.
The 3$\times$3 matrices ${\cal H}_0({\bf k})$ and $L_z$ are given by

\begin{equation}
{\cal H}_0 ({\bf k})=
\begin{bmatrix}
V_0 & V_1 & V_2 
\\
V^{*}_1 & V_{11} & V_{12}
\\
V^{*}_2 & V^{*}_{12} & V_{22}
\end{bmatrix}
,~~~~
L_z=
\begin{bmatrix}
0 & 0 & 0 
\\
0 & 0 & -i
\\
0 & i & 0
\end{bmatrix} 
\label{eq:1b}
\end{equation}

There are more details regarding the matrix elements of ${\cal H}_0 ({\bf k})$; see supplementary information in He. {\it et al}~\cite{He:2016}.
The band structure of ML-NS in the normal state, obtained by a tight binding calculation, shown in Fig.~\ref{fig:1}(c).
Since the number of electrons in the outermost of an Nb atom is one less than in Mo and W atoms,
the chemical potential of ML-NS, denoted with a green dashed-line in Fig.~\ref{fig:1}(c), lies inside the valence band. 
Noted that the dominated orbitals in the valence-band of ML-NS are the $d_{z^2}$,~$d_{xy}$ and $d_{x^2-y^2}$ orbitals of the Nb atoms~\cite{Lebegue:2009,Liu:2013}.
As shown in Fig.~\ref{fig:1}(c), the strong Ising SOC leads to enhanced spin-split as large as 150meV~\cite{Radisavljevic:2011} in the valence bands near the K points.
As a result, it is worth mentioning that the superconducting states near the K points survive in the presence of a strong in-plane magnetic field.
In ML-NS, in addition to the two Fermi pockets at the corners
of the Brillouin zone (BZ) with opposite crystal momentum ($\pm$K) similar to other TMDs,
there is an extra pocket around the $\Gamma$ point of BZ at the Fermi level. 
However, very weak spin splitting valence bands for the spin-up and spin-down electrons around the $\Gamma$
point could lead to an alternative strategy for unconventional pairing possibilities
between the $d_{z^2}$ orbitals, which are dominant around the $\Gamma$ point~\cite{Radisavljevic:2011}.
The Bogoliubov-de Gennes Hamiltonian of ML-NS including spin-singlet pairing reads as follows~\cite{He:2016}:

 \begin{align}
 \begin{aligned}
&
{\cal H}_{\rm ML-NS}\!=\!\sum_{\bf k}  {\bm \Omega}^\dagger_{{\bf k}}
{\cal H'({\bf k})} {\bm \Omega}_{{\bf k}},
\label{eq:2}
 \end{aligned}
 \end{align}
 where ${\bm \Omega}^{\dagger}_{\bf k}=( c^{\dagger}_{{\bf k}\uparrow}, c^{\dagger}_{{\bf k}\downarrow},
c_{-{\bf k}\uparrow}, c_{-{\bf k}\downarrow})$ is the Nambu space operator and
 \begin{align}
 \begin{aligned}
&
{\cal H'({\bf k})}\!=\!E_{\bf k} ({\tau}_z\!\otimes\!{\sigma}_0)
+\lambda_{so}\vartheta_{\bf k} ({\tau}_0\!\otimes\!{\sigma}_z)+ \Delta_{sc} ({\tau}_y\!\otimes\!{\sigma}_y),
\\
&
E_{\bf k}= \frac{\hbar ^2 {k}^2}{2m_{\rm v}}-\mu_{\rm v}, \quad  \quad \vartheta_{\bf k}={k}_+^3+{k}_-^3 ,
\label{eq:2a}
 \end{aligned}
 \end{align}
in which the parameters $m_\mathrm{v}$ and $\mu_{\mathrm{v} }$ correspond to the valence-band effective mass and the Fermi energy measured from top of the valence band, respectively.
The superconducting energy gap of ML-NS, defined by $\Delta_{\rm sc}$ and $k=\sqrt{k_x^2+k_y^2}$, is the magnitude of the in-plane momentum.
Here, $\tau_i$ and $\sigma_i$ ($i=x,y,z$) are Pauli matrices in the particle-hole and spin spaces, respectively.
Moreover, $\tau_0$ is corresponding to an identity matrix.
Here, we set $\lambda_{so}$=80 meV and $\Delta_{sc}$=5 meV.

To explore the induced-superconducting in an ML-NS due to the ferromagnetic proximity effect,
we consider a heterostructure including a ML-NS and a ferromagnet substrate as schematically
shown in Fig.~\ref{fig:1}(a).
The spin polarized electrons of a ferromagnet substrate in Nambu space can be expressed at
the basis ${\bm \varphi}_{\bf q}=(a_{{\bf q}\uparrow}, a_{{\bf q}\downarrow}, a^{\dagger}_{-{\bf q}\uparrow},a^{\dagger}_{-{\bf q}\downarrow})$ as follows:

\begin{equation}
 {\cal H}_{\rm F}=\sum_{\bf q}{\bm \varphi}^{\dagger}_{\bf q}
 \begin{pmatrix}
{ H}_{\rm F}({\bf q})& 0 \\
0 &  -{ H}_{\rm F}^*(-{\bf q})
\end{pmatrix} {\bm \varphi}_{\bf q} \quad,
 \label{eq:3}
\end{equation}
where
\begin{equation}
{ H}_{\rm F}({\bf q})= \varepsilon_{\bf q} 
\sigma_0  - {\bf J} \cdot \boldsymbol{\sigma},~~~ \\
\varepsilon_{\bf q} = \hbar^2 q^2/2m_e- \mu_{\rm F}
 \label{eq:3a}
\end{equation}
in which ${\bf J}$ is an exchange field with components as ${\bf J}=(J_x,J_y,J_z)$ and $q=|{\bf q}|$.   
Here, $a^{\dagger}_{{\bf q}\sigma}~(a_{{\bf q}\sigma})$ operator creates (annihilates) an electron with momentum vector ${\bf q}=(q_x,q_y)$
and spin $\sigma=\uparrow,\downarrow$ inside the ferromagnet substrate.
The parameters $m_{\rm e}$ and $\mu_{\rm F}$ address electron mass and Fermi energy in ferromagnet substrate, respectively.  
To investigate the possibility of realizing unconventional SC in ML-NS due to the scattering of Cooper pairs from the ferromagnet substrate,
one can assume a tunneling between the ferromagnet substrate and ML-NS, which can be modeled via a coupling Hamiltonian as follows: 

\begin{align}
 \begin{aligned}
 &
{\cal H}_{\rm coupling} = \sum_{{\bf q},{\bf k}} 
{\bm \Omega}^{\dagger}_{\bf k} {\cal H}_{\rm T} {\bm \varphi}_{\bf q}  +{\rm H.c.},
\\
&
 {\cal H}_{\rm T} = t ({\tau}_z \otimes{\sigma}_0).
\label{eq:4}
 \end{aligned}
 \end{align}
in which the parameter $t$ is the tunnel coupling between ML-NS and the ferromagnet substrate in the junction
with the same spin and momentum.
The total Hamiltonian for describing the ML-NS/ferromagnet heterostructure can be defined as

\begin{equation}
 {\cal H}_{total}={\cal H}_{\rm ML-NS} + {\cal H}_{\rm F} + {\cal H}_{\rm coupling}.
\label{eq:5}
\end{equation}

 
\section{formalism}

We proceed to inspect the symmetry of the modified Cooper pairs in ML-NS due to the ferromagnetic proximity effect 
with an arbitrary magnetization direction with respect to the Ising SOC.
For this purpose, we evaluate the modified pairing correlations in
ML-NS by utilizing the Matsubara Green's function formalism~\cite{Bruus:2004}.
The uncoupled Green's-function of ML-NS is written as:  

\begin{align}
 \begin{aligned}
&
G^{0}_{\rm ML-NS}({\bf k},i\omega_n)=
\begin{pmatrix}
 \hat{g}({\bf k},i\omega_n) &  \hat{f}({\bf k},i\omega_n)  \\
\hat{f}^\dag({\bf k},i\omega_n) & -\hat{g} (-{\bf k},-i\omega_n)
\end{pmatrix},\\[0.1cm]
\\
&
\hat{g}({\bf k},i\omega_n) = \frac{(\xi_{\bf k,\nu}+i\omega_n){\sigma}_0 }{(i\omega_n)^2 -
\left( \xi_{\bf k,\nu}^2+\Delta_{\rm sc}^2 \right) }
\\
&
 \hat{f} ({\bf k},i \omega) = \frac{\nu \Delta_{\rm SC}}{(i\omega_n)^2 - \left( \xi_{{\bf k},\nu}^2+\Delta_{\rm sc}^2 \right)},
\label{eq:6}
 \end{aligned}
 \end{align}
where $\xi_{{\bf k},\nu}=E_{\bf k}+\nu\lambda_{so} \vartheta_{\bf k}$  is the
electron excitation spectrum with subband index $\nu=\pm1$ standing for the spin-up and spin-down energy bands.
The Green's function of the ferromagnet substrate can be expressed as $G^{0}_{\rm F}({\bf q},i\omega_n)=[i\omega_n -\mathcal{H}_{\rm F} (\bf {q})]^{-1}$,
which is a block diagonal matrix in the Nambu spinor representation as follows: 
\begin{equation}
G^{0}_{\rm F}({\bf q},i\omega_n)=
\begin{pmatrix}
\frac{\left(\varepsilon_{\bf q} -i \omega_n\right){\sigma}_0  + {\bf J} \cdot \boldsymbol{\sigma}}{|{\bf J}|^2-\left( \varepsilon_{\bf q} -i \omega_n \right)^2} & 0\\
0 & -\frac{\left(\varepsilon_{-\bf q} +i \omega_n\right){\sigma}_0  + {\bf J} \cdot {\boldsymbol \sigma}^{\star}}{| {\bf J}|^2-\left( \varepsilon_{-\bf q} +i \omega_n \right)^2}
\label{eq:7}
\end{pmatrix}
\end{equation}

Using Dyson equation, the total Green's function of the whole model Hamiltonian ${\cal H}_{{\rm ML-NS}-{\rm F}}$ reads as follows: 
 \begin{align}
 \begin{aligned}
 &
\hspace{-.05cm}{G}_{\rm {ML-NS}}({\bf k},i\omega_n)=G^0_{\rm {ML-NS}}({\bf k},i\omega_n)
\\
&
\hspace{1.5cm} +G^{0}_{\rm {ML-NS}}({\bf k},i\omega_n)\Sigma(\mathbf{k},i\omega_n)\,{G}_{\rm {ML-NS}}({\bf k},i\omega_n).
\label{eq:8}
  \end{aligned}
 \end{align}
The magnetic proximity effect on ML-NS comes into play through the self-energy, which has the following form:
\begin{equation}
\Sigma({\bf k},i\omega_n)= \sum_{\bf q} {\cal H}_{\rm T}
G^{0}_{\rm F}({\bf q},i\omega_n) {\cal H}_{\rm T}^\dag.
\label{eq:8a}
\end{equation}
The Green's function of ML-NS due to the ferromagnetic proximity effect using Eq.~(\ref{eq:8}) is given by: 
\begin{align}
 \begin{aligned}
 {G}_{\rm ML-NS}({\bf k},i\omega_n) = \left[ (G^{0}_{\rm ML-NS}({\bf k},i\omega_n))^{-1} - \Sigma({\bf k},i\omega_n) \right]^{-1}.
 \label{eq:9}
\end{aligned}
 \end{align}

In the following, we will focus on the weak coupling regime where the ferromagnetic proximity effects minimally influence the electronic band structure of the ML-NS. 
Therefore, it is convenient to approximate the total Green's function in the second-order perturbation
of the tunneling parameter $t$ as the following expression:

\begin{align}
 \begin{aligned}
  &
\hspace{-.05cm} {G}_{\rm ML-NS}({\bf k},i\omega_n) \approx G^{0}_{\rm ML-NS}({\bf k},i\omega_n)
\\
&
\hspace{2.cm} + G^{0}_{\rm ML-NS}({\bf k},i\omega_n)\Sigma({\bf k},i\omega_n) G^{0}_{\rm ML-NS}({\bf k},i\omega_n).
\label{eq:10}
 \end{aligned}
 \end{align}

As a result, the modified anomalous Green's functions of ML-NS are given by:
 \begin{align}
 \begin{aligned}
\hspace{0.cm}  {F}_{\nu\nu}({\bf k},i\omega_n )= \frac{t^2 \nu \hat{f} ({\bf k},i\omega_n ) ( \nu' J_x+i J_y )}{(i\omega_n)^2 - ( \xi_{{\bf k},\nu'}^2+\Delta_{\mathrm{SC}}^2 )}\times
   \\
   &
\hspace{-4.5cm} \left[\frac{ \xi_{{\bf k},\nu}+i \omega_n}{|{\rm{\bf J}}|^2-(-\varepsilon_{\bf q}+i\omega_n)^2}- \frac{ \xi_{{\bf k},\nu'}-i \omega_n}{|{\rm{\bf J}}|^2-(\varepsilon_{\bf q}+i\omega_n)^2} \right]
 \\
 &
\hspace{-6.3cm}
{F}_{\nu\nu'}({\bf k},i\omega_n )= \hat{f}({\bf k},i\omega_n ) + \frac{ t^2\hat{f}({\bf k},i\omega_n )}{ (i\omega_n)^2-( \xi_{{\bf k},\nu'}^2+\Delta_{\mathrm{SC}}^2 )}\times
 \\
 &
\hspace{-6.7cm}  \left[ \frac{(\xi_{{\bf k},\nu}+i\omega_n)(\varepsilon_{\bf q}+\nu J_z-i\omega_n)}{|{\rm{\bf J}}|^2-(\varepsilon_{\bf q}-i\omega_n)^2}+
  \frac{(\xi_{{\bf k},\nu}-i\omega_n)(\varepsilon_{\bf q}-\nu J_z+i\omega_n)}{|{\rm{\bf J}}|^2-(\varepsilon_{\bf q}+i\omega_n)^2}  \right] 
 \label{eq:11}
 \end{aligned}
 \end{align}

It should also be noted that an electron with momentum ${\bf k}$ can only tunnel to a state in the ferromagnet substrate with momentum ${\bf q}={\bf k}$
due to the momentum conservation~\cite{Triola2014},
and assume any spin-flip scattering process occurs at the ML-NS/ferromagnet heterostructure.  
 
\section{symmetry of the induced superconducting correlations}

At zero temperature, the retarded anomalous Green's function is afforded by substituting the Matsubara frequency by its analytic continuation
$i\omega_n\rightarrow \omega + i\eta $.
Hereinafter, for simplicity we ignore the imaginary part, i.e., $\eta$.
To investigate the symmetry of the induced superconducting pairing correlations, we rewrite the retarded anomalous Green's function in a matrix form as follows:  
\begin{equation}
{\hat{\mathcal{F}}}({\bf k},\omega)= [ {\mathcal{F}}^s({\bf k},\omega) \sigma_0 + {\bm{\mathcal{F}}}^t({\bf k},\omega)\cdot \boldsymbol{{\bf \sigma}} ] i\sigma_y,
\label{eq:12}
\end{equation} 
in which the singlet ${\bf\mathcal{F}}^s({\bf k},\omega)$ and triplet ${\bm{\mathcal{F}}}^t({\bf k},\omega)$ pairing correlations
can be obtained using the modified anomalous Green's functions in Eq.~(\ref{eq:11}) as follows:

\begin{align}
 \begin{aligned}
 &
 {\bf\mathcal{F}}^s({\bf k},\omega)=\frac{1}{2}(F_{\uparrow\downarrow}-F_{\downarrow\uparrow}),~~~~{\mathcal{F}}_x^t({\bf k},\omega)=\frac{1}{2}(F_{\downarrow\downarrow}-F_{\uparrow\uparrow}),
 \\
 &
 {\mathcal{F}}_y^t({\bf k},\omega)=\frac{-i}{2}(F_{\uparrow\uparrow}+F_{\downarrow\downarrow}),
 ~~~~{\mathcal{F}}_z^t({\bf k},\omega)=(F_{\uparrow\downarrow}+F_{\downarrow\uparrow}).
 \label{eq:13a}
 \end{aligned}
\end{align}

The induced singlet and triplet pairing correlations in ML-NS at zero temperature in more details are given by
\begin{widetext}
\begin{align}
 \begin{aligned}
{\mathcal{F}}^{\rm{s}}({{\bf k},\omega)} &= -\frac{4t^2 {\hat{f}}^{2}({\bf k},\omega)}{ \mathcal{L}({\bf k},\omega) (\omega^2- \xi_{{\bf k},\nu}^2 - \Delta_{\mathrm{SC}}^2 ) (\omega^2 - \xi_{{\bf k},\nu'}^2 - \Delta_{\mathrm{SC}}^2 )} \left(\mathcal{Z}({\bf k},\omega)-2J_z\lambda_{so}\omega\vartheta_{\bf k}\mathcal{Z}'({\bf k},\omega)\right),
 \\
   &
\hspace{-1.40cm} {\mathcal{F}}^{\rm{t}}_{x,y}({\bf k},\omega) =-\frac{2t^2 {\hat{f}}^{2}({\bf k},\omega)}{\mathcal{L}({\bf k},\omega)}\left({\bf\mathcal{F}}^{\rm{even}}_{x,y}({\bf k},\omega) + {\bf\mathcal{F}}^{\rm{odd}}_{x,y}({\bf k},\omega) \right),
\\
&
\hspace{-1.40cm} {\mathcal{F}}^{\rm{even}}_{x,y}({\bf k},\omega) =i \left({\bf J}\times \hat{z}\right) \lambda_{so}\vartheta_{\bf k} \left( -\varepsilon_{\bf k} ^2+|{\bf J}|^2-\omega^2 \right),
\hspace{1.50cm} {\mathcal{F}}^{\rm{odd}}_{x,y} ({\bf k},\omega) =  J_{x,y} \omega \left( \varepsilon_{\bf k}^2 + 2\varepsilon_{\bf k} E_{\bf k} - |{\bf J}|^2 + \omega^2 \right),
\\
&
\hspace{-1.40cm} {\mathcal{F}}^{\rm{t}}_{z} ({\bf k},\omega) = \frac{4t^2  \hat{f}^{2}({\bf k},\omega)}{\mathcal{L}({\bf k},\omega) (\omega^2 - \xi_{{\bf k},\nu}^2 - \Delta_{\mathrm{SC}}^2 ) (\omega^2- \xi_{{\bf k},\nu'}^2 - \Delta_{\mathrm{SC}}^2 )}   \left(\lambda_{so} \vartheta_{\bf k} \mathcal{K}({\bf k},\omega) + J_z\omega\mathcal{K}'({\bf k},\omega) \right),
\label{eq:13}
\end{aligned}
 \end{align}
 \end{widetext} 
in which $\mathcal{L}({\bf k},\omega)$,~$\mathcal{Z}({\bf k},\omega)$,~$\mathcal{Z}^{'}({\bf k}$,$\omega),~\mathcal{K}({\bf k}$,$\omega)$
and $ \mathcal{K}'({\bf k},\omega)$ are even functions in frequency (see the Appendix).
To be consistent with Fermi-Dirac statistics,
there are different classes of Cooper pairs based on the possible symmetries of the pairing correlations in three degrees of freedom, .i.e., frequency, spin and spatial parity.
As is clear from Eq.~(\ref{eq:13}), the unique Ising SOC supports an
admixture state of a spin-singlet state and an opposite spin-triplet state ($m=0$)
in the absence of the induced exchange field.
Production of the opposite spin-triplet pairing correlation originates from the fact that the triplet pairing should be aligned to the SOC direction,
${\bm{\mathcal{F}}}^t({\bf k},\omega)||{\hat z}$~\cite{Frigeri:2004aa}.
In other words, the equal-spin spin-triplet pairing correlations $(m=\pm1)$ are exactly zero
because ${\cal H}_{\rm ML-NS}$ commutes with $\sigma_z$.
In the absence of the exchange field, the admixture state is fully even
in frequency due to the time-reversal symmetry,
although our results show that the novel admixture state including both odd- and even-frequency components
for when the induced exchange field is collinear with respect to the Ising SOC.
While the induced magnetization is misaligned with the spin quantization axis,
the interplay of SOC and the induced magnetization results in a noncollinearity ferromagnet at the Fermi level, as illustrated schematically in Fig.~\ref{fig:1}(d).
As a result, the existence of the equal spin-triplet pairing correlations
guaranteed in an ML-NS owing to the desirable effect of the misaligning magnetization.
Unlike equal spin-triplet pairing correlations that have been demonstrated to be in plane in the multi-layer systems due to the non-collinear ferromagnet,
the equal spin-triplet pairings in ML-NS aligned out of plane ~\cite{Halterman2007,Halterman2008,Kawabata2013}.

\section{results}

To proceed to classify the pairing correlations according to their symmetry properties with respect to three degrees of freedom, i.e., frequency, spin, and spatial parity,
we define the renormalized pairing correlations by the singlet gap parameter in the absence of the proximity effect:

\begin{align}
 \begin{aligned}
 &
 {\bar{\mathcal{F}}}^{\rm{s}}({\bf k},\omega)={\mathcal{F}}^{\rm{s}}({\bf k},\omega)/{\mathcal{F}}^{\rm{s}}_0,
\\
&
 {\bm{\bar{\mathcal{F}}}}^{\rm{t}}({\bf k},\omega)={\bm{\mathcal{F}}}^{\rm{t}}({\bf k},\omega)/{\mathcal{F}}^{\rm{s}}_0.
 \end{aligned}
 \end{align}

At first sight, let us start to study the symmetry properties of the pairing correlations with respect to frequency.
The even-frequency superconducting order parameter is characterized as ${\bar{\mathcal{F}}}({\bf k},\omega=0)$.
Since ${\bm{\bar{\mathcal{F}}}}^{\rm{t}}({\bf k},\omega)$ always diminishes at equal time ($\omega=0$)
as required by the Pauli principle, the odd-frequency superconducting order parameter is defined as ${{\partial_{\omega}{\bar{\mathcal{F}}}}({\bf k},\omega)|_{\omega=0}}$.
Magnitudes of the even- and odd-frequency superconducting pairings
in ML-NS for an arbitrary magnetization $\bf J$ in the $xz$ plane are shown in Fig.~\ref{fig:2}.
Note that we choose a special ${\bf k}$ in our calculations; however, the same qualitative behavior is obtained for other values.
In the absence of the induced magnetization, 
an admixture state of spin-singlet and opposite spin triplet pairings is generated in ML-NS that is even in frequency.
According to our results, the odd-frequency order parameter begins to grow by increasing the induced magnetization
irrespective of its direction and the even-frequency order parameter decreases by simultaneous measuring.
In order to favor the {\it odd-frequency spin-triplet even-parity} order parameter, the induced exchange field needs to be out of plane.
The importance of the {\it odd-frequency spin-triplet even-parity} pairing correlation is more closely related to the existence of the Majorana fermions
~\cite{Tanaka:2012,Asano:2013,Snelder:2015,Huang:2015,Ebisu:2015,Kashuba:2017}.
The even-parity nature of this special pairing plays a very important role in robustness against disorder~\cite{Eschrig2008}. 

\begin{figure}[ht]
\vspace{-0.7cm}
\includegraphics[width=0.85 \linewidth]{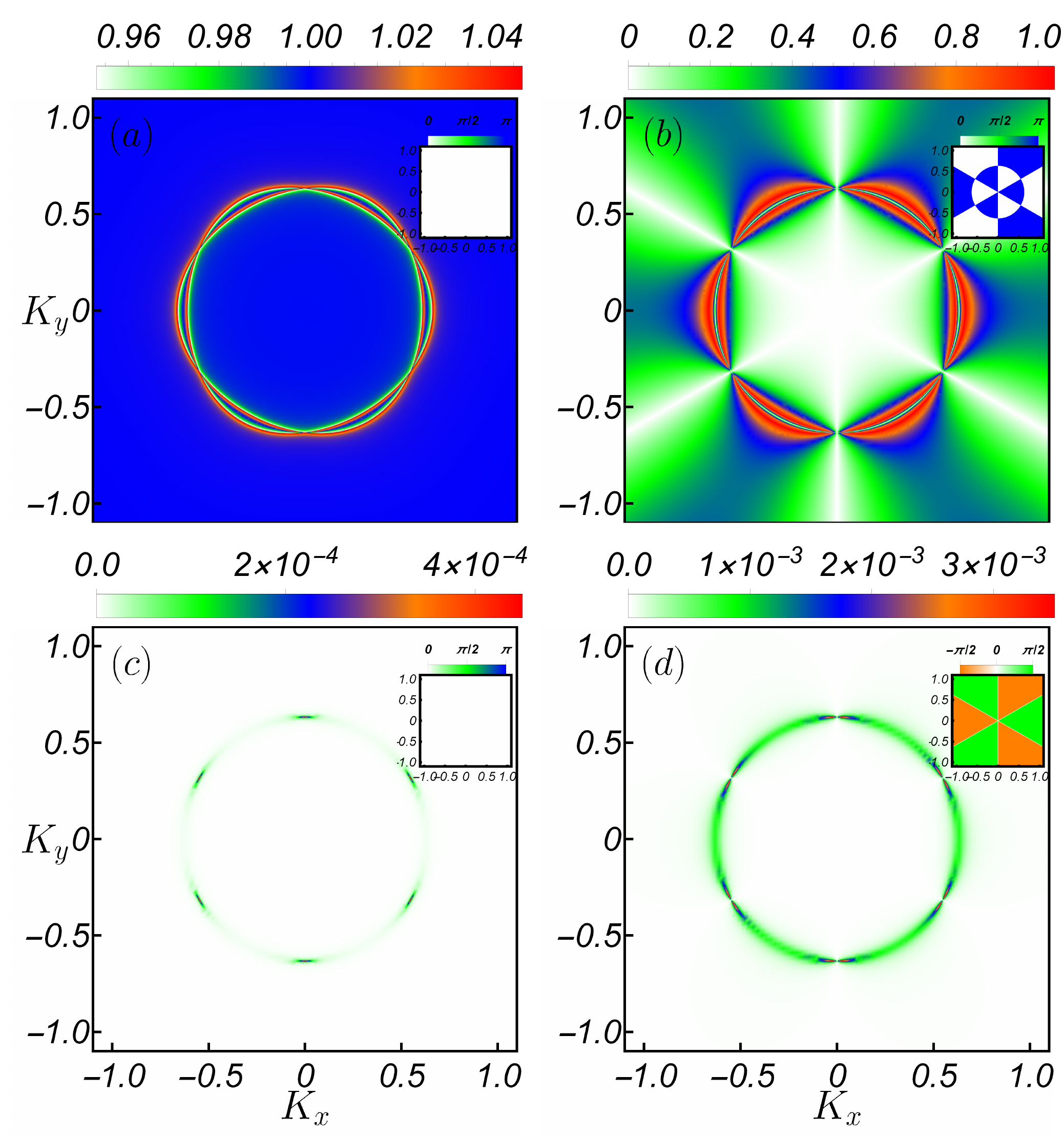}
\vspace{-0.1cm}
\caption{
The $k$ dependence of amplitudes and complex phases (insets) of the modified pairing correlations,
(a) $\bar{\bf\mathcal{F}}^{\rm{s}}({\bf k},\omega)$,
(b) $\bar{\bf\mathcal{F}}^{\rm{t}}_{z}({\bf k},\omega)$, (c) $\bar{\bf\mathcal{F}}^{\rm{t}}_{x}({\bf k},\omega)$, and (d) $\bar{\bf\mathcal{F}}^{\rm{t}}_{y}({\bf k},\omega)$,
for a finite magnetization $J$=1.5 eV in the $x$ direction and ${\mu_{\rm F}}$=12 eV.
The results shown are for $\omega$=2 meV, but the behavior is the same for all frequencies as well.
} 
\label{fig:3}
\end{figure}

Next, we investigate the parity symmetry of the pairing correlations
for a finite magnetization to be oriented in the $x$ direction ($\theta=\pi/2$ and $\phi=0$).
As is apparent in Eq.~(\ref{eq:13}), the $x$ component of the equal spin-triplet pairing is odd in frequency
while ${\bf\mathcal{F}}^{\rm{s}}({\bf k},\omega)$, ${\bf\mathcal{F}}^{\rm{t}}_{z}({\bf k},\omega)$, and  ${\bf\mathcal{F}}^{\rm{t}}_{y}({\bf k},\omega)$ 
are even in frequency.
To illustrate the momentum dependencies of the anomalous Green's functions,
we plot amplitudes and complex phases of the anomalous Green's functions in Fig.~\ref{fig:3}.
Our results are entirely consistent with the Pauli principle.
Since the complex phases of 
${\bf\mathcal{F}}^{\rm{s}}({\bf k},\omega)$ and  ${\bf\mathcal{F}}^{\rm{t}}_{x}({\bf k},\omega)$
are unchanged for any ${\bf k}$, these components are even in parity.
While for the other anomalous Green's functions, ${\bf\mathcal{F}}^{\rm{t}}_{z}({\bf k},\omega)$, and  ${\bf\mathcal{F}}^{\rm{t}}_{y}({\bf k},\omega)$,
the $\pi$ phase change is always present by changing the sign of ${\bf k}$, then these pairing correlations are odd in parity.

It is worth noting that the spin singlet and the opposite spin-triplet pairing correlations become dominant to
the equal spin-triplet components.
The spin-singlet pairing has a $s$-wave symmetry state that is a fully gapped superconducting state,
while the opposite spin-triplet pairing acquires six nodal lines in the corners of the BZ; therefore, it has a $f$-wave symmetry state. 
As shown in Fig.~\ref{fig:3}(b), the complex phase of ${\bf\mathcal{F}}^{\rm{t}}_{z}({\bf k},\omega)$ near the BZ corners changes suddenly, causing
the opposite spin triplet pairing to stay odd in parity.
As a result, the interpocket pairing state allowed on the pocket at $\Gamma$ point owing to
both the $s$-wave and $f$-wave pair states belongs the same irreducible representation of the point group $D_{3h}$.

\begin{figure}
\vspace{-0.75cm}
\includegraphics[width=1.015 \linewidth]{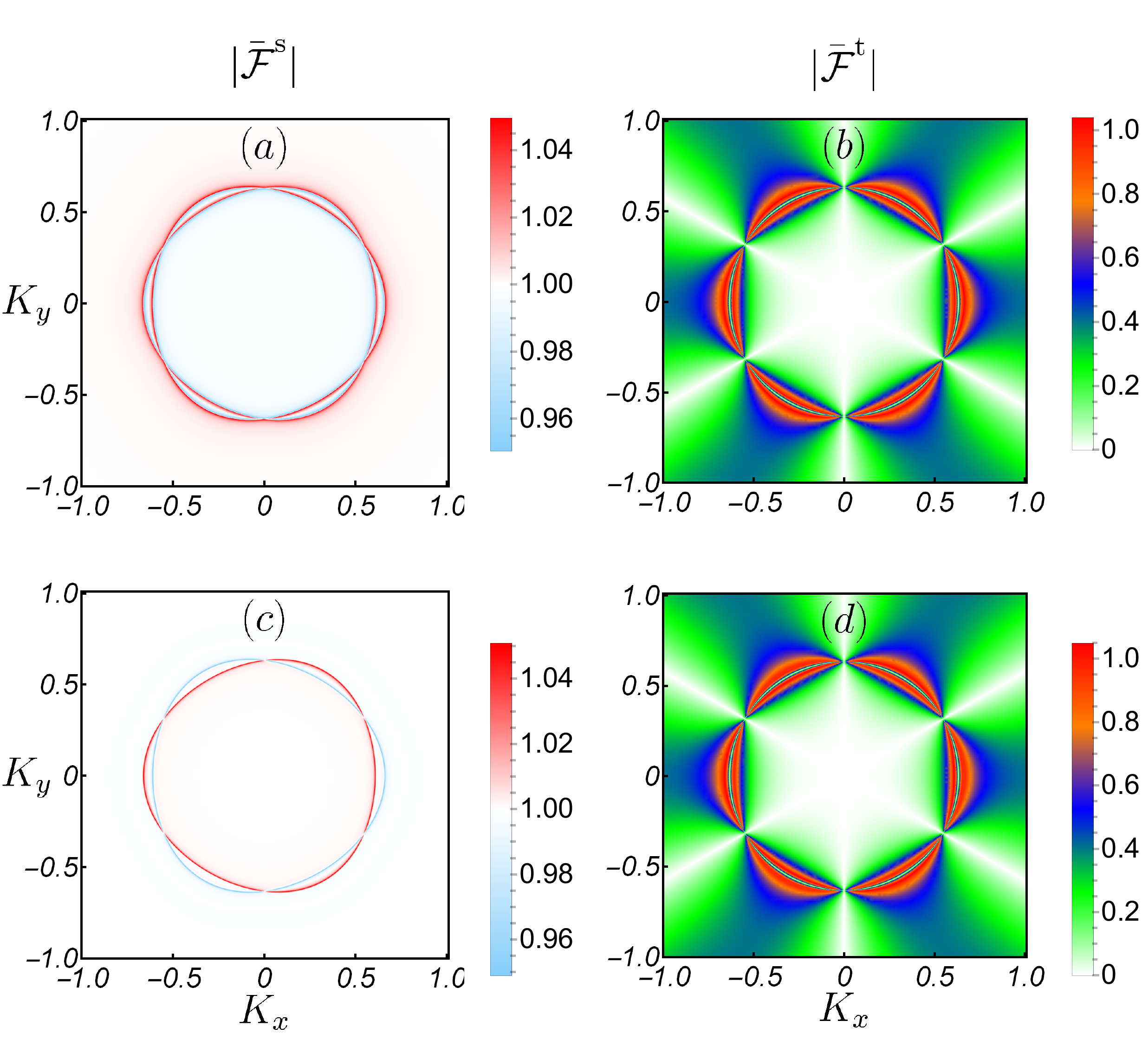}
\vspace{-0.7cm}
\caption{
Amplitudes of $\bar{\bf\mathcal{F}}^{\rm{s}}({\bf k},\omega)$ (left)
$\bar{\bf\mathcal{F}}^{\rm{t}}({\bf k},\omega)$ and (right) induced in ML-NS
when the external magnetization is ${\bf J}=(J_x,0,0)$.
The results shown are for the metallic ferromagnet substrate in the top panels with $J$=1.5 eV,$\mu_{\rm F}$=12 eV and the half-metallic ferromagnet substrate in the bottom panels
with $J$=10 eV and $\mu_{\rm F}$=2 eV.
The results shown are for $\omega$=2 meV, with the same behavior for all frequencies as well.
} 
\label{fig:4}
 \end{figure}

Next, we present in Fig.~\ref{fig:4} the pairing amplitudes of the induced admixture state in ML-NS
due to the magnetic proximity effect of different magnetic substrates.
Note that the ratio of the exchange field to the Fermi energy is smaller than 1
for the metallic ferromagnet, while the ratio of $|\mathbf{J}|/\mu_F$ is larger than 1 for a fully spin-polarized half-metallic ferromagnet.
Cobalt, for example, is a metallic ferromagnet with $|\mathbf{J}|/\mu_F\approx0.1$~\cite{Cedex1975}
and CrO$_2$ is a half metallic ferromagnet with $|\mathbf{J}|/\mu_F\approx1$~\cite{Singh:2015}.
As is apparent in Fig.~\ref{fig:4}, the proximity effect of a half-metallic ferromagnet is seen to result in
a reduction in the magnitude of the spin-singlet pairing correlation related to the metallic ferromagnet substrate.
By considering the spin quantization axis along the $z$ direction,
the spin state of an electron at ML-NS ($|\alpha\rangle$) for when the magnetization  has only in plane components, can be expressed as a superposition 
of the spin states of the ferromagnet substrate ($|\pm\rangle$), as $|\alpha \rangle= n |+\rangle+m|-\rangle$.
Therefore, electrons with the up- and down-spins in ML-NS can tunnel through, into, and out of the ferromagnet substrate, while for another spin, the ferromagnet substrate would be an insulator. It is worth mentioning that only the phase of the electron is completely reﬂected from the ferromagnet interface, be affected by scattering process

To understand the difference in the magnitude of the spin singlet pairing correlation in ML-NS due to the proximity effect originated
for the various ferromagnet substrates,
we define the wave function of the spin singlet Cooper pairs at the proximity of a ferromagnet substrate as:
$\psi^{s}\propto |r_{\nu}||r_{\nu'}|e^{(\eta+\eta')} \left( |\uparrow\downarrow\rangle - |\downarrow\uparrow\rangle \right)$,
in which $r_{\nu}$ and $\eta$ are the reflection amplitude and the phase shift due to the scattering of electrons from the ferromagnet interface.
In the case of the metallic ferromagnet, both spin reflection amplitudes are smaller than one ($|r_{\nu,\nu'}|<1$)
while the reflection amplitude of one kind of spin is perfectly equal to one in the half-metallic ferromagnet. Therefore, this causes the Cooper pair wave function to be more intensely affected in the case of the half-metallic substrate.

\section{conclusions}
Experimental studies have demonstrated the existence of an Ising SC in ML-NS
in an in-plane critical upper magnetic field several times higher than the Pauli paramagnetic limit. 
In ML-NS, in addition to the two Fermi pockets at the valley points ($\pm$K) similar to other TMDs, there is an extra
pocket around the $\Gamma$ point of BZ at Fermi level with weak spin splitting valence bands for the spin-up and
spin-down electrons around the $\Gamma$ point, which could lead to an alternative strategy for new pairing possibilities.
For this purpose, we started from the Bogoliubov-de Gennes Hamiltonian of
ML-NS including spin singlet pairing 
to investigate the induced unconventional superconducting phase in ML-NS due to the ferromagnet proximity effect.
In order to characterize the symmetry of the induced pairing correlations in ML-NS,
we obtained the anomalous Green's-functions via the Matsubara Green's function formalism.
Our analytical results clearly show that a mixed singlet-triplet superconductivity, $s+f$,
can be generated in ML-NS, that includes both odd- and even-frequency components
due to the ferromagnetic proximity effect.
Frequency-symmetry of the pairing correlations can be manipulated by changing the magnitude and direction of the induced magnetization.
There exists the possibility to extend our results for describing the induced unconventional superconducting state
in a monolayer TaS$_2$
owing to the fact that monolayer TaS$_2$ has the similar electronic structure and the same symmetry properties as ML-NS, but with the stronger Ising SOC~\cite{Barrera:2018}. 
As a result, the upper critical field in monolayer TaS$_2$ is remarkably increased
due to the large intrinsic SOC.  
%


\appendix

\section{Functions Defining the Anomalous Green's-Function}
\label{appA}
The functions introducing the modified singlet and triplet pairing correlations, Eq.~(\ref{eq:13}), are defined as

\begin{widetext}
\begin{align}
 \begin{aligned}
 \mathcal{L}({\bf k},\omega) &= \Delta_{\mathrm{SC}}\left(|{\bf J}|^2-(\varepsilon_{\bf k} -\omega)^2 \right) \left(|{\bf J}|^2 + (\varepsilon_{\bf k} +\omega)^2 \right)~,  
 \\
 &
\hspace{-1.1cm} \mathcal{Z}({\bf k},\omega) = \varepsilon_\bk E_{\bk}\left( \varepsilon_\bk^2-|{\bf J}|^2-\omega^2  \right)\mathcal{X}({\bf k},\omega)+\omega^2\left( \varepsilon_\bk^2+|{\bf J}|^2-\omega^2  \right)\mathcal{Y}({\bf k},\omega) ~,
\\
&
\hspace{-1.1cm} \mathcal{Z}'({\bf k},\omega) = 2 E_\bk\left(- \varepsilon_\bk^2+|{\bf J}|^2-\omega^2  \right) \mathcal{M}({\bf k},\omega) -\varepsilon_\bk \mathcal{N}({\bf k},\omega) ~,
\\
&
\hspace{-1.1cm} \mathcal{K}({\bf k},\omega) = 4E_\bk\omega^2\left( \varepsilon_\bk^2+|{\bf J}|^2+\omega^2  \right)\mathcal{M}({\bf k},\omega)+ \left( \varepsilon_\bk^3+\varepsilon_\bk\left(|{\bf J}|^2+\omega^2 \right) \right) \mathcal{N}({\bf k},\omega)
\\
&
\hspace{-1.1cm} \mathcal{K}'({\bf k},\omega) = -2\varepsilon_\bk E_\bk \mathcal{X}({\bf k},\omega)+\left( -\varepsilon_\bk^2+|{\bf J}|^2-\omega^2  \right)\mathcal{Y}({\bf k},\omega) ~,
\\
&
\hspace{-1.1cm} \mathcal{X}({\bf k},\omega) = E_\bk ^4+\left( 2E_\bk ^2 + \Delta_{\mathrm{SC}}^2-3 \lambda_{so}^2 \vartheta_\bk ^2 -\omega ^2 \right)\left( \Delta_{\mathrm{SC}}^2+\lambda_{so}^2 \vartheta_\bk^2-\omega^2 \right),
\\
&
\hspace{-1.1cm} \mathcal{Y}({\bf k},\omega) = E_\bk ^4+\left( \Delta_{\mathrm{SC}}^2+\lambda_{so}^2 \vartheta_\bk ^2-\omega^2 \right)^2+2E_\bk ^2\left( \Delta_{\mathrm{SC}}^2+3\lambda_{so}^2 \vartheta_\bk ^2 -\omega ^2 \right),
\\
&
\hspace{-1.1cm} \mathcal{M}({\bf k},\omega) = E_\bk ^2 + \Delta_{\mathrm{SC}}^2 + \lambda_{so}^2 \vartheta_\bk ^2 -\omega ^2,
\\
&
\hspace{-1.1cm} \mathcal{N}({\bf k},\omega) =   3E_\bk ^4-\left( \Delta_{\mathrm{SC}}^2+\lambda_{so}^2 \vartheta_\bk ^2-\omega^2 \right)^2 + 2E_\bk ^2\left( \Delta_{\mathrm{SC}}^2-\lambda_{so}^2 \vartheta_\bk ^2 -\omega ^2 \right). \nn\\
\end{aligned}
 \end{align}
 \end{widetext} 
 
\bibliography{References}
\end{document}